\documentclass[11pt]{article}

\usepackage[table]{xcolor}

\usepackage[preprint]{acl}
\usepackage{times}
\usepackage{latexsym}
\usepackage{placeins}
\usepackage[T1]{fontenc}

\usepackage[utf8]{inputenc}

\usepackage{microtype}

\usepackage{inconsolata}

\usepackage{graphicx}

\usepackage{booktabs}
\usepackage{multirow}
\usepackage{amsmath}
\usepackage{amssymb}
\usepackage{tcolorbox}
\usepackage{svg}
\title{Listen, Reason, and Segment: Aligning LALMs with Editorial Judgment for Media Chapterization}

\author{
  Tony Alex%
  \thanks{Corresponding authors: \texttt{t.alex@surrey.ac.uk} and
  \texttt{j.deng16@imperial.ac.uk}.}%
  \thanks{This work was undertaken during Tony Alex's internship at MVP Lab, London.} \\
  University of Surrey
  \And
  Wish Suharitdamrong \\
  University of Surrey
  \And
  Sara Atito \\
  University of Surrey
  \AND
  Armin Mustafa \\
  University of Surrey
  \And
  Muhammad Awais \\
  University of Surrey
  \And
  Philip J. B. Jackson \\
  University of Surrey
  \AND
  Jiankang Deng\footnotemark[1] \\
  MVP Lab
  \And
  Ismail Elezi \\
  MVP Lab
}

\begin{document}
\maketitle
\begin{abstract}
Large Audio Language Models (LALMs) have made rapid progress on standardized benchmarks, yet their deployment in practical media workflows, curation, archival indexing, and content distribution remains largely unrealized. We identify automated audio chapterization, the task of segmenting continuous audio streams into thematically coherent chapters, as a demanding and commercially consequential setting that exposes this gap. Chapterization is challenging because boundaries are defined less by objective acoustic events than by subjective editorial judgment, requiring models to reason sequentially over long acoustic contexts and approximate creator-authored boundary decisions. We present \textbf{AudioChaps}, a post-training framework for aligning end-to-end LALMs for this task via Group Relative Policy Optimization (GRPO) guided by Chain-of-Thought (CoT) reasoning. To support training and evaluation, we curate three datasets: \textit{AudioChaps-Alignment}, derived from creator-annotated chapter boundaries on YouTube; \textit{AudioChaps-CoT}, which provides structured supervision for well-formatted, high-quality, and evidence-grounded boundary reasoning; and \textit{AudioChaps-Eval}, a held-out benchmark for audio chapterization. Applying GRPO directly without a Supervised Fine-Tuning (SFT) cold start, \textit{AudioChaps-R1-Zero} already improves average F1 by 33 points over the state-of-the-art LALM Audio-Flamingo-3-Think. The AudioChaps framework produces our final aligned LALM, \textit{AudioChaps-R1}, which improves average F1 by 49 points. These results demonstrate that GRPO-trained LALMs can reliably transform unstructured auditory streams into navigable, structured media. Our code, models, and dataset resources will be released upon acceptance at \url{https://github.com/ta012/AudioChaps}.
\end{abstract}

\section{Introduction}
\label{label_intro}

Multimodal Large Language Models (MLLMs), spanning Vision-Language Models (VLMs)~\citep{zhu2023minigpt,tong2024cambrian}, Large Audio Language Models (LALMs)~\citep{deshmukh2023pengi,gong2024listen,tang2024salmonn,alex2026palprobingaudioencoders,chu2024qwen2audiotechnicalreport, ghosh2025audio}, and their variants, have rapidly advanced domain-specific understanding across modalities ~\cite{thawkar2023xraygpt, maaz2023video, brohan2023rt}. Within the auditory domain in particular, LALMs have evolved through a now-standard recipe of large-scale pretraining, Supervised Fine-Tuning (SFT), and Reinforcement Learning from Human Feedback (RLHF) \citep{tian2025stepaudior1technicalreport}, yielding models with increasingly nuanced acoustic comprehension.

However, strong performance on benchmark datasets~\citep{sakshi2024mmau,ma2025mmar,wang2025mmsu} is only a proxy for the broader ambition of \emph{intelligence as a utility}: the point at which these models are embedded in real-world workflows and demonstrably improve quality and efficiency. Realizing this ambition requires moving beyond generalized capability evaluations toward targeted, task-specific deployments grounded in concrete application domains.

\begin{figure*}[t]
    \centering
    \includegraphics[width=0.78\textwidth]{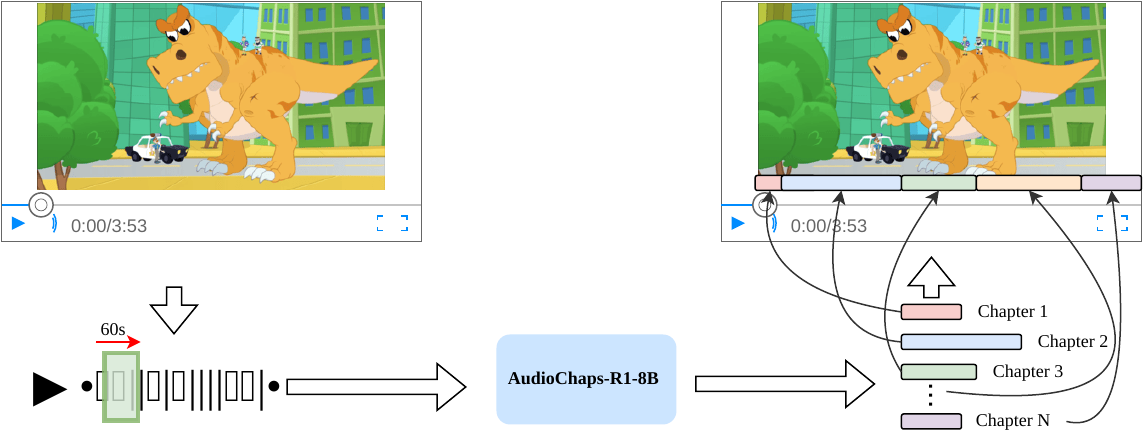}
    \caption{\small Illustration of the audio chapterization task: the media's audio is analysed in 60-s chunks via AudioChaps-R1-8B to determine the chapter boundaries, presented to the user as a timeline above the scrubber bar.}
    \label{fig_main_diag}
\end{figure*}


The media and entertainment industry offers a particularly compelling testbed for this transformation. The sector has a persistent and growing need for AI-assisted tools, both in the production of new content and in the repurposing of archival catalogues for modern distribution platforms. Concretely, this includes editorial assistance (cut suggestion, subtitling, shot generation), structured metadata creation for indexing and retrieval, and chapterization for navigable consumption. The scale and economic stakes are substantial: legacy broadcasters such as CNN and the BBC must curate decades of archival footage for streaming re-release, while platforms including Netflix, YouTube, and Spotify rely on accurate structured metadata to make heterogeneous content libraries efficiently discoverable.

Among these tasks, audio chapterization stands out as both broadly applicable and methodologically demanding: it requires reasoning over continuous acoustic streams to recover the latent thematic structure that human editors would impose. Given an audio stream of arbitrary duration, the system must partition it into thematically coherent chapters. While narrow speech-only settings such as podcasts can be partially addressed by cascading Automatic Speech Recognition (ASR) with a text-LLM, such pipelines degrade sharply on the heterogeneous audio that defines real-world media: dynamic editorial content, music, gaming streams, and beyond. We therefore target an end-to-end LALM that operates directly on raw audio and generalizes across these regimes.

The importance of this structuring extends well beyond efficiency. Web search transformed an unstructured collection of pages into a navigable medium and, in doing so, became the substrate on which much of the modern web economy was built. Fine-grained media navigation promises a comparable shift for audio and video: once long-form content is segmented into semantically coherent units, a further layer of services becomes possible, from precise retrieval and recommendation to personalised education and training, with downstream implications for business, technical upskilling, wellbeing, and healthcare. Chapterization is a foundational step toward this navigable layer, turning continuous streams into addressable structure on which higher-level applications can build on.

A central difficulty is that chapter boundaries are defined less by objective acoustic events than by subjective editorial judgment. Models must therefore reason over distributed acoustic evidence and learn a decision rule that reflects how creators actually structure content. Because uniformly imitating synthetic reasoning traces may produce plausible explanations without yielding well-calibrated boundary decisions, we use GRPO~\citep{shao2024deepseekmathpushinglimitsmathematical, Guo_2025} to optimize the model's final decisions against creator-authored chapter annotations. In our primary instantiation, AudioChaps-R1 is trained in two stages: SFT on the AudioChaps-CoT corpus establishes a structured, evidence-grounded reasoning prior, after which GRPO calibrates boundary decisions toward creator-authored annotations.

Our primary contributions are as follows:
\begin{itemize}
    \item We propose \textbf{AudioChaps}, a post-training framework for aligning LALMs for audio chapterization. It uses GRPO to optimize boundary decisions against creator-authored editorial annotations, and the resulting aligned LALM is termed \textbf{AudioChaps-R1}.

    \item We introduce \textbf{AudioChaps-Alignment}, a chapterization dataset derived from creator-annotated YouTube content and \textbf{stratified across four acoustic regimes} (structured speech, dynamic media, gaming, and music), moving the task beyond the speech-only setting to which existing transcript-based pipelines are limited.

    \item We present \textbf{AudioChaps-CoT}, a reasoning corpus constructed through a novel \textbf{audio-to-text modality bridge}. It provides structured supervision for well-formatted, high-quality, and evidence-grounded boundary reasoning, addressing the absence of supervised reasoning data for subjective audio chapterization.

    \item We release \textbf{AudioChaps-Eval}, the first benchmark dedicated to \textbf{audio-only chapterization}, enabling principled comparison on a task with no prior standardized evaluation.

    \item Through extensive experiments, AudioChaps-R1 establishes \textbf{state-of-the-art performance across all four acoustic regimes}, raising average F1 from 28.6 to 77.8 over its base model and surpassing a 32B RL-trained LALM at roughly a quarter of the parameters.
\end{itemize}

\section{Related Works}

\noindent \textbf{Audio and Video Chapterization.}
Chapterization has been studied almost entirely in the video domain. VidChapters-7M~\citep{yang2023vidchapters7mvideochaptersscale} introduced a large corpus of user-chaptered YouTube videos and the tasks of chapter generation, generation given ground-truth boundaries, and grounding. More recent systems extend to hour-long content with richer supervision, combining ASR transcripts and visual tokens under instruction tuning~\citep{pu2025arcchapterstructuringhourlongvideos}. Deployed tools likewise remain transcript-driven, running a text-LLM over ASR output, which confines them to speech-dominant material. To the best of our knowledge, no prior work targets chapterization from audio alone. We close this gap with an end-to-end model that operates on raw audio, exploiting non-speech cues such as music and ambient transitions, and that treats boundary placement as editor-supervised acoustic reasoning rather
than transcript segmentation, allowing it to generalize across speech, music, gaming, and dynamic media.

\section{Methodology}
\label{sec_methodology}

In this section, we detail the development and training of AudioChaps using AF3-Think-8B~\citep{goel2025audio}, a strong contemporary LALM, as our primary backbone. We begin by formally defining the audio chapterization task (Section~\ref{sec_task_definition}). Next, we evaluate the zero-shot chapterization performance of AF3-Think-8B to establish the baseline for measuring the improvements introduced by AudioChaps (Section~\ref{sec_baseline}). We then investigate whether GRPO can directly align the unmodified AF3-Think-8B model with creator-authored chapter annotations (Section~\ref{sec_challenge_rl}). Finally, we present our primary two-stage AudioChaps instantiation, which combines a CoT Supervised Fine-Tuning (SFT) cold start with GRPO-based calibration (Section~\ref{sec_audiochap_framework}).

\subsection{Task Formulation}
\label{sec_task_definition}

We formulate audio chapterization as boundary detection over short audio segments. For training and evaluation, we sample positive and negative 60-second clips from each source video. For positive clips, the ground-truth chapter boundary is sampled uniformly within the central 20 seconds of the window, between 20 and 40 seconds, ensuring at least 20 seconds of acoustic context before and after the transition. Negative clips lie strictly within a single chapter and are sampled with a temporal buffer from any annotated boundary. For each clip, the model is tasked with determining whether a thematic transition is present. This construction provides sufficient pre- and post-transition context for the model to reason over narrative flow rather than rely on instantaneous acoustic cues. At deployment, the same model can be applied using a sliding window over audio streams of arbitrary duration (Figure~\ref{fig_main_diag}; Supplementary Section~\ref{apx_continuous_eval}), enabling chapterization without requiring native long-context audio modelling, which remains an open challenge for current LALMs.

Ground-truth boundaries are sourced from creator-annotated chapter markers on YouTube, which serve as a naturally occurring proxy for human editorial judgment (Section~\ref{sec_dataset}). A limitation of AF3-Think-8B is that it does not emit boundary timestamps with sufficient accuracy to support reliable temporal reasoning supervision (Section~\ref{sec_cot_data_generation}). We therefore formulate the primary training objective as a flow-based, presence-or-absence judgment, keeping both the reasoning traces and final decisions grounded in signals that can be supervised reliably.

\subsection{How Much Can AudioChaps Improve an Existing LALM?}
\label{sec_baseline}

We use Audio-Flamingo-3-Think-8B (AF3-Think-8B)~\citep{goel2025audio} as the primary backbone throughout our main experiments and first evaluate its zero-shot chapterization performance to establish the unaligned starting point. AF3-Think is the on-demand reasoning variant of Audio Flamingo 3 and does not enforce an explicit reasoning format. Its zero-shot performance therefore provides the principal reference point for quantifying the improvements introduced by AudioChaps.

We additionally evaluate Step-Audio-R1-32B~\citep{tian2025stepaudior1technicalreport}, a substantially larger audio-reasoning model used to generate the acoustic perception logs during AudioChaps-CoT curation. Its inclusion enables us to assess whether the final AudioChaps-R1 model reflects only the capabilities of a model involved in its supervision pipeline or, through task-specific alignment, develops stronger chapterization ability. Broader comparisons with general-purpose multimodal and omni-modal models are provided in Supplementary Section~\ref{apx_comparision_large_mllm}.

\subsection{The Challenge of Direct RL: AudioChaps-R1-Zero}
\label{sec_challenge_rl}

Having established the zero-shot chapterization performance of AF3-Think-8B, we next ask whether reinforcement learning alone can align its decisions with the editorial judgment reflected in creator-authored annotations. Recent work on text-only reasoning models has shown that GRPO-style optimization with rule-based rewards can elicit substantial reasoning capabilities from unaligned base models, an approach popularized as ``R1-Zero''~\citep{Guo_2025}. We investigate the analogous setting for audio by applying GRPO directly to AF3-Think-8B, yielding \textit{AudioChaps-R1-Zero}.

A direct port of the R1-Zero recipe is, however, impractical in our setting. AF3-Think-8B does not natively emit responses in the strict \texttt{<think>...</think><answer>...</answer>} schema; enforcing this format via reward would penalize nearly all initial rollouts, collapsing the learning signal. We therefore adapt the reward function to the model's native output style: rather than requiring exact tag-based formatting, we reward (i) the presence of a discernible reasoning trace preceding the final decision and (ii) the correctness of the binary boundary verdict. This formulation preserves the spirit of rule-based RL while remaining tractable for an unaligned starting point.

While direct GRPO yields measurable gains over the AF3-Think-8B baseline, qualitative inspection reveals inconsistent reasoning styles and weak grounding in task-relevant acoustic evidence across regimes. For our primary AF3-based instantiation, we therefore introduce a CoT cold start before GRPO. This stage standardizes the output format and provides an evidence-grounded reasoning initialization, after which GRPO calibrates boundary decisions against creator-authored annotations.

\subsection{The AudioChaps Framework}
\label{sec_audiochap_framework}

This section presents the primary instantiation of AudioChaps on AF3-Think-8B. We first describe the construction of \textit{AudioChaps-CoT} through an audio-to-text modality bridge (Section~\ref{sec_cot_data_generation}), and then detail the SFT cold start and GRPO calibration used to obtain AudioChaps-R1 (Section~\ref{sec_sft_grpo}).

\subsubsection{CoT Data Generation for Audio Chapterization}
\label{sec_cot_data_generation}

\begin{figure*}[t]
    \centering
    \includegraphics[width=0.78\textwidth]{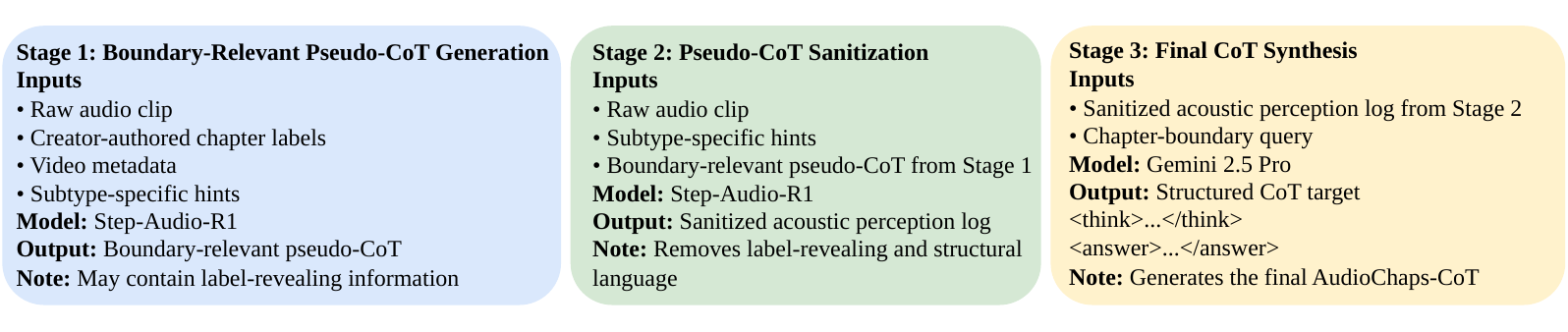}
    \caption{\small
    AudioChaps-CoT curation pipeline. A boundary-relevant pseudo-CoT is generated, sanitized into an acoustic perception log, and then converted into the final structured CoT target used for the SFT cold start.
    }
    \label{fig_cot}
\end{figure*}

Directly applying GRPO requires the model to discover both the desired reasoning strategy and output format from sparse binary rewards. To provide a structured initialization, we construct \textit{AudioChaps-CoT} through an audio-to-text modality bridge (Figure~\ref{fig_cot}). The pipeline converts continuous audio into a sanitized acoustic perception log that a stronger text-based reasoning model can use to generate evidence-grounded CoT targets.

\paragraph{Stage 1: Pseudo-CoT Generation.}
For each clip, Step-Audio-R1 generates an initial pseudo-CoT conditioned on the audio, subtype, video title, and creator-authored chapter supervision. Positive samples are grounded using the semantic change between adjacent chapter titles, while negative samples are identified as continuous segments. This supervision encourages the model to identify task-relevant acoustic evidence rather than produce a generic audio description.

\paragraph{Stage 2: Sanitized Acoustic Perception Log.}
Because the pseudo-CoT contains explicit label and metadata supervision, it cannot be used directly for SFT. Step-Audio-R1 therefore processes the audio again, using the pseudo-CoT only as contextual guidance, and produces a chronological description of audible evidence. Explicit verdicts and structural terms such as ``boundary,'' ``transition,'' ``segment,'' and ``chapter'' are removed, yielding a perception log that preserves relevant acoustic cues without revealing the target label.

\paragraph{Stage 3: Final CoT Synthesis.}
Gemini 2.5 Pro~\citep{comanici2025gemini25pushingfrontier} receives only the sanitized perception log and the binary chapterization query. It infers whether a boundary is present and generates the final target in the required \texttt{<think>...</think><answer>...</answer>} format. The synthetic CoT targets describe the evolution of the acoustic evidence using flow-based language rather than exact timestamps. This design is particularly suitable for LALMs such as AF3-Think-8B, which we found did not predict temporal locations consistently enough to support reliable timestamp-based supervision. The resulting corpus provides well-formatted, evidence-grounded targets for the SFT cold start.

\subsubsection{SFT Cold Start and GRPO Calibration}
\label{sec_sft_grpo}

\begin{figure*}[t]
\small
\begin{equation}
\label{eq_grpo}
\begin{aligned}
J_{\mathrm{GRPO}}(\theta)
&=
\mathbb{E}_{
q \sim P(Q),\;
\{o_i\}_{i=1}^{G} \sim \pi_{\theta_{\mathrm{old}}}(O \mid q)
}
\\[-0.4em]
&\quad
\left[
\frac{1}{G}\sum_{i=1}^{G}
\min\left(
\frac{\pi_{\theta}(o_i \mid q)}
     {\pi_{\theta_{\mathrm{old}}}(o_i \mid q)}
A_i,\;
\mathrm{clip}\left(
\frac{\pi_{\theta}(o_i \mid q)}
     {\pi_{\theta_{\mathrm{old}}}(o_i \mid q)},
1-\epsilon,
1+\epsilon
\right)A_i
\right)
-
\beta D_{\mathrm{KL}}\left(
\pi_{\theta}\parallel\pi_{\mathrm{ref}}
\right)
\right].
\end{aligned}
\end{equation}
\end{figure*}

Equipped with the \textit{AudioChaps-CoT} corpus, we align AF3-Think-8B through a two-stage procedure: an SFT cold start followed by GRPO-based calibration.

\paragraph{Stage 1: Cold-Start SFT.}
AF3-Think-8B does not reliably produce well-formatted, high-quality reasoning traces for boundary detection. We therefore fine-tune it on \textit{AudioChaps-CoT} (prompt template in Supplementary Section~\ref{apx_train_prompt}) to obtain \textit{AudioChaps-SFT}. This stage establishes the structured output format required for rule-based GRPO rewards, consistently separating the reasoning trace from the final verdict, while providing a coherent initialization for subsequent calibration.

\paragraph{Stage 2: GRPO-Based Calibration.}
Although SFT provides a structured initialization, it encourages the model to imitate a single synthetic rationale for each example. This can produce plausible, well-formatted explanations without adequately calibrating when a boundary should be predicted. We therefore apply GRPO to \textit{AudioChaps-SFT}, sampling a group of $G$ rollouts for each query, where each rollout may follow a different reasoning trajectory before producing a boundary verdict. GRPO assigns higher relative advantage to trajectories whose final decisions agree with creator-authored chapter annotations, thereby calibrating the model's boundary judgments. Unlike SFT, which learns from one prescribed rationale, GRPO can reinforce multiple reasoning trajectories that lead to the same creator-aligned decision. This flexibility is well suited to the subjective nature of chapterization, where the same editorial boundary may be supported by different combinations of semantic, discourse, acoustic, and structural cues.

Formally, as shown in Equation~\ref{eq_grpo}, we optimize the policy $\pi_{\theta}$ over trajectories $\{o_i\}_{i=1}^{G}$ sampled conditional on the audio query $q$. Here, $\pi_{\theta_{\mathrm{old}}}$ denotes the policy used to generate the current rollout group, while $\pi_{\mathrm{ref}}$ is the frozen \textit{AudioChaps-SFT} reference policy. The clipped surrogate objective bounds the policy ratio within $[1-\epsilon,1+\epsilon]$, and the Kullback--Leibler penalty $\beta D_{\mathrm{KL}}(\pi_{\theta}\parallel\pi_{\mathrm{ref}})$ limits divergence from the SFT initialization.

For each rollout $o_i$, we compute an unweighted scalar rule-based reward:
\begin{equation}
\label{eq_reward}
r_i =
r_{\mathrm{format}}(o_i)
+
r_{\mathrm{accuracy}}(o_i),
\end{equation}
where $r_{\mathrm{format}}$ verifies the presence and validity of the \texttt{<think>} and \texttt{<answer>} tags, and $r_{\mathrm{accuracy}}$ measures agreement between the predicted boundary decision and the creator-authored label. Both components are binary, taking value $1$ when the corresponding criterion is satisfied and $0$ otherwise.

The group-relative advantage is obtained by normalizing each reward within the $G$ rollouts sampled for the same query:
\begin{equation}
\label{eq_advantage}
A_i =
\frac{r_i-\mu_r}{\sigma_r+\delta},
\qquad
\mu_r =
\frac{1}{G}\sum_{j=1}^{G}r_j,
\end{equation}
where $\sigma_r$ is the standard deviation of the group rewards and $\delta$ is a small constant for numerical stability. A rollout therefore receives positive advantage when its reward exceeds that of the other reasoning trajectories sampled for the same audio query. This rule-based formulation enforces valid structured outputs while directly aligning boundary decisions with editorial supervision, without requiring a learned reward model.

Overall, SFT supplies the structured reasoning interface required for stable optimization, while GRPO provides the decisive alignment signal that calibrates boundary judgments to creator-authored editorial supervision.

\begin{table}[t]
\centering
\small
\begin{tabular}{p{18.0ex}lcccc}
\toprule
{\mbox{ }\hspace{-1.5ex}\textbf{Model}} & \textbf{sc} & \textbf{Acc} & \textbf{Pre} & \textbf{Rec} & \textbf{F1} \\
\midrule
& DM      & 60.2 & 55.3 & 62.7 & 58.8 \\
{\mbox{ }\hspace{-1.5ex}Step-Audio-R1-32B\hspace{-2ex}\mbox{ }}
& G             & 60.1 & 59.1 & 53.4 & 56.1 \\
{\mbox{ }\hspace{-1.5ex}(Zero-Shot)}
& M              & 62.4 & 53.1 & 51.9 & 52.5 \\
& SS  & 65.6 & 70.1 & 69.7 & 69.9 \\
\cmidrule(l){2-6}
&\textit{Avg}& 62.1 & 59.4 & 59.4 & 59.3 \\
\midrule
& DM       & 55.8 & 47.5 & 23.7 & 31.6 \\
{\mbox{ }\hspace{-1.5ex}AF3-Think-8B}
& G             & 50.1 & 45.9 & 54.6 & 49.9 \\
{\mbox{ }\hspace{-1.5ex}(Zero-Shot)}
& M              & 62.3 & 64.0 &  3.2 &  6.0 \\
& SS  & 44.7 & 46.4 & 19.1 & 27.0 \\
\cmidrule(l){2-6}
&\textit{Avg}& 53.2 & 51.0 & 25.2 & 28.6 \\
\midrule
& DM       & \textbf{75.9} & \textbf{70.0} & \textbf{77.2} & \textbf{73.4} \\
{\mbox{ }\hspace{-1.5ex}AudioChaps-R1-8B\hspace{-2ex}\mbox{ }}
& G             & \textbf{77.6} & \textbf{75.1} & \textbf{75.9} & \textbf{75.5} \\
{\mbox{ }\hspace{-1.5ex}(Ours)}
 & M              & \textbf{88.1} & \textbf{83.6} & \textbf{85.6} & \textbf{84.6} \\
& SS  & \textbf{73.6} & \textbf{71.0} & \textbf{86.0} & \textbf{77.8} \\
\cmidrule(l){2-6}
&\textit{Avg}& \textbf{78.8} & \textbf{74.9} & \textbf{81.2} & \textbf{77.8} \\
\bottomrule
\end{tabular}
\caption{
Performance comparison on AudioChaps-Eval across the four audio
chapterization subcategories: \textit{Dynamic Media} (DM),
\textit{Gaming} (G), \textit{Music} (M), and
\textit{Structured Speech} (SS). Accuracy (Acc), precision (Pre),
recall (Rec), and F1 are reported as percentages, together with the
macro-average (Avg) over subcategories. Detailed comparison with additional multimodal and omni-modal LLMs is provided in Supplementary Table~\ref{tab_advanced_lalm_comparison}.
}
\label{table_main_results}
\end{table}

\section{Experiments}
\subsection{Dataset}
\label{sec_dataset}

\paragraph{Data Source and Acoustic Stratification.}
We curate our training and evaluation data from \textit{VidChapters-7M}~\citep{yang2023vidchapters7mvideochaptersscale}, a large-scale collection of YouTube videos with chapter boundaries manually provided by content creators. To evaluate audio chapterization across heterogeneous media, we query the YouTube Data API for each video's category label and group the selected content into four broad acoustic subtypes: \textit{structured speech} (Education, Science \& Technology, and Howto \& Style), \textit{dynamic media} (Entertainment, Comedy, and Film \& Animation), \textit{gaming}, and \textit{music}. This stratification exposes the model to diverse acoustic conditions, ranging from narrated instructional content to edited entertainment, gameplay audio, and music-dominant recordings.

\paragraph{Clip Construction.}
For each source video, we construct positive and negative 60-second clips. A positive clip contains a creator-authored chapter boundary, with the boundary sampled uniformly between 20 and 40 seconds of the clip. This placement provides at least 20 seconds of acoustic context before and after the annotated transition. Negative clips are sampled entirely within a single annotated chapter and are kept temporally separated from annotated boundaries. Where possible, negative clips are sampled from the same source videos as positive clips, reducing differences in speakers, recording conditions, and production style between the two classes.

\paragraph{Training Data.}
The resulting \textit{AudioChaps-Alignment} training set contains approximately 30k labelled clips across the four acoustic subtypes, including 13,347 positive boundary-spanning clips and 16,636 negative within-chapter clips. A 22k-clip subset is used to construct \textit{AudioChaps-CoT} for the SFT cold start, while the labelled alignment data provides creator-authored supervision for GRPO.

\paragraph{Clip-Level Evaluation.}
For evaluation, we construct \textit{AudioChaps-Eval}, a held-out test set containing approximately 16k clips from 749 source videos, including 7,011 positive clips and 8,952 negative clips. The training and test partitions are split at the source-video level, so no source video appears in both sets. Detailed per-subtype counts and class distributions are provided in Supplementary Section~\ref{apx_data_stats}.

\paragraph{Full-Length Evaluation.}
In addition to the controlled clip-level \textit{AudioChaps-Eval} benchmark, we evaluate chapter-boundary detection on a balanced set of full-length recordings spanning all four acoustic subtypes. The set comprises 40 recordings, with 10 from each subtype, and contains 387 reference boundaries. The recordings have a mean duration of 49 minutes, yielding approximately 33 hours of long-form audio in total. The continuous-stream inference and evaluation protocol is described in Section~\ref{sec_results} and Supplementary Section~\ref{apx_continuous_eval}.

\subsection{Results Discussion}
\label{sec_results}

Table~\ref{table_main_results} compares two zero-shot baselines with our final aligned model on \textit{AudioChaps-Eval}. AF3-Think-8B~\citep{goel2025audio} serves as the backbone for AudioChaps-R1, while Step-Audio-R1-32B~\citep{tian2025stepaudior1technicalreport} is used to generate the acoustic perception logs during AudioChaps-CoT curation. Importantly, the boundary labels are provided by neither model and remain grounded in creator-authored annotations. Comparing against both models therefore evaluates the gains over the underlying backbone as well as over the larger model involved in the supervision pipeline. Additional comparisons with multimodal and omni-modal LLMs are provided in Supplementary Table~\ref{tab_advanced_lalm_comparison}.

\paragraph{Zero-Shot LALMs Find Chapterization Difficult.}
AF3-Think-8B exhibits a conservative prediction pattern: on \textit{Music}, it achieves 64.0 precision but only 3.2 recall, indicating that it predominantly predicts ``no boundary'' and identifies only the clearest transitions. The same tendency appears in \textit{Structured Speech} and \textit{Dynamic Media}, where recall remains below 25. Step-Audio-R1-32B produces a more balanced precision--recall profile, but its accuracy remains between 60\% and 65\% across all four subtypes, indicating that zero-shot chapterization remains challenging even for a substantially larger RL-trained LALM.

\paragraph{Comparison with an ASR--LLM Cascade.}
To test whether chapter boundaries can be identified from transcripts alone, we construct a cascade baseline using Whisper-Large-V3~\citep{radford2022robustspeechrecognitionlargescale} for automatic speech recognition(ASR), followed by Qwen3-235B-A22B-Instruct-2507-FP8~\citep{yang2025qwen3} to analyse the transcript and determine whether a chapter boundary is present. We evaluate this baseline on the \textit{Structured Speech} subset, where ASR is expected to be most reliable. The cascade achieves 71.4 precision, 36.1 recall, and 48.0 F1, compared with 71.0 precision, 86.0 recall, and 77.8 F1 for AudioChaps-R1. The comparable precision but substantially lower recall suggests that transcripts capture some explicit semantic transitions but miss boundaries signalled by acoustic pacing, delivery changes, pauses, and background continuity.

\paragraph{AudioChaps-R1 outperforms both its backbone LALM and the larger LALM used for CoT curation.}
AudioChaps-R1 yields substantial and consistent gains over the AF3-Think-8B backbone. Averaged across subtypes, accuracy rises from 53.2 to 78.8, precision from 51.0 to 74.9, recall from 25.2 to 81.2, and F1 from 28.6 to 77.8. A paired video-level bootstrap analysis with 10,000 resamples confirms that AudioChaps-R1 improves macro-F1 over AF3-Think-8B by 49.2 points, with a 95\% confidence interval of [46.2, 52.2] and a two-sided bootstrap p-value below $10^{-4}$. Most of the improvement comes from recall, reflecting the backbone's conservative tendency to predict no boundary. The largest gain occurs on \textit{Music}, where recall rises from 3.2 to 85.6 and F1 from 6.0 to 84.6. These recall gains do not come at the expense of precision, which also increases across all subtypes, including from 47.5 to 70.0 on \textit{Dynamic Media} and from 46.4 to 71.0 on \textit{Structured Speech}. The largest accuracy gain occurs on \textit{Structured Speech}, increasing from 44.7 to 73.6. AudioChaps-R1 also outperforms Step-Audio-R1-32B, the larger LALM used to produce the acoustic perception logs during CoT curation, by 14.6, 19.4, 32.1, and 7.9 F1 points on \textit{Dynamic Media}, \textit{Gaming}, \textit{Music}, and \textit{Structured Speech}, respectively, despite using approximately one quarter of its parameters. Its improvement over Step-Audio-R1-32B indicates that AudioChaps-R1 does not merely inherit the capabilities of a model involved in its supervision pipeline, but develops stronger chapterization ability through task-specific alignment.

\paragraph{AudioChaps-R1 performs consistently across acoustic regimes.}
AF3-Think-8B exhibits substantial variation across subtypes, spanning 43.9 F1 points from 6.0 on \textit{Music} to 49.9 on \textit{Gaming}. In contrast, AudioChaps-R1 narrows this range to 11.2 points, from 73.4 on \textit{Dynamic Media} to 84.6 on \textit{Music}. This more consistent performance indicates that the aligned model remains effective across diverse acoustic regimes rather than favouring a single content type.

\paragraph{Full-length chapter-boundary detection.}
For full-length evaluation, we apply the model using a sliding 60-second window with a 20-second hop (Figure~\ref{fig_main_diag}). Consecutive windows predicted as containing a boundary are grouped into a positive run, from which a single boundary estimate is placed at the temporal midpoint.

As shown in Table~\ref{tab:continuous_eval}, the zero-shot AF3-Think-8B backbone achieves an F1 score of 6.5 on uncropped, full-length recordings. A fixed-interval baseline that places a boundary every 180 seconds reaches 9.5 F1, whereas AudioChaps-R1-8B increases F1 to 37.6 and reduces the pooled median reference-to-estimate deviation from 38.0 to 10.0 seconds.

These results show that local boundary decisions can be consolidated into effective recording-level chapter predictions, providing a practical approach to long-form audio chapterization without relying on native long-context audio modelling, which remains limited in current LALMs. Additional details are provided in Supplementary Section~\ref{apx_continuous_eval}.

\begin{table}[t]
\centering
\small
\begin{tabular}{lrrrr}
\toprule
\textbf{Model} &
\textbf{F1} &
\textbf{Pre} &
\textbf{Rec} &
\textbf{dev-R2E} \\
\midrule
Fixed interval, 180 s
& 9.5 & 9.0 & 11.6 & 49.0 s \\
AF3-Think-8B
& 6.5 & 8.6 & 9.6 & 38.0 s \\
\textbf{AudioChaps-R1-8B}
& \textbf{37.6} & \textbf{34.1} &
\textbf{45.7} & \textbf{10.0 s} \\
\bottomrule
\end{tabular}
\caption{Full-length chapter-boundary detection on uncropped audio. Predictions are matched one-to-one with reference boundaries under a $\pm 10$-second tolerance. Precision, recall, and F1 are computed per recording, macro-averaged within each acoustic subtype, and then averaged equally across the four subtypes. \mbox{dev-R2E} denotes the pooled median temporal distance from each reference boundary to its nearest predicted boundary. Detailed results are provided in Supplementary Table~\ref{tab_long_audio_full}.}
\label{tab:continuous_eval}
\end{table}

\subsection{Ablation Study}
\label{sec_ablation}

\begin{table}[t]
\centering
\small
\begin{tabular}{lrrrr}
\toprule
\textbf{Model} & \textbf{DM} & \textbf{G} & \textbf{M} & \textbf{SS} \\
\midrule
AF3-Think-8B & 31.6 & 49.9 & 6.0 & 27.0 \\
AudioChaps-R1-Zero-8B & 57.8 & 64.4 & 59.1 & 62.9 \\
AudioChaps-SFT-8B & 70.0 & 72.5 & 74.1 & \textbf{77.9} \\
\textbf{AudioChaps-R1-8B} & \textbf{73.4} & \textbf{75.5} & \textbf{84.6} & 77.8 \\
\bottomrule
\end{tabular}
\caption{Ablation of the AudioChaps training stages. F1 scores are reported for \textit{Dynamic Media} (DM), \textit{Gaming} (G), \textit{Music} (M), and \textit{Structured Speech} (SS). Detailed results are provided in Supplementary Section~\ref{apx_detailed_ablation}, Table~\ref{tab_detailed_ablation}.}
\label{tab:ablation_f1}
\end{table}

Table~\ref{tab:ablation_f1} isolates the contribution of each training stage by comparing the AF3-Think-8B backbone, direct GRPO without an SFT cold start, SFT alone, and the full SFT+GRPO pipeline. F1 is reported across \textit{Dynamic Media}, \textit{Gaming}, \textit{Music}, and \textit{Structured Speech}, with full per-metric results provided in Supplementary Section~\ref{apx_detailed_ablation}, Table~\ref{tab_detailed_ablation}. All variants use AF3-Think-8B as the underlying backbone.

\paragraph{Direct RL yields substantial but uneven gains.}
Applying GRPO directly to the AF3-Think-8B backbone produces AudioChaps-R1-Zero-8B, raising F1 to 57.8, 64.4, 59.1, and 62.9 on \textit{Dynamic Media}, \textit{Gaming}, \textit{Music}, and \textit{Structured Speech}, respectively. These correspond to gains of 26.2, 14.5, 53.1, and 35.9 points over the backbone. Much of the improvement is driven by increased recall, indicating that direct GRPO reduces the backbone's tendency to default to ``no boundary.'' Notably, this recall-oriented shift resembles the trend produced by SFT, but remains consistently weaker across all four subtypes. This suggests that, without a structured reasoning prior, GRPO exploration primarily shifts the model toward greater boundary sensitivity rather than producing a sufficiently well-calibrated chapterization policy. These results support the use of a CoT cold start before applying GRPO to AF3-Think-8B.

\paragraph{SFT cold-start establishes a high-recall prior.}
AudioChaps-SFT-8B improves over AudioChaps-R1-Zero across all four subtypes, reaching F1 scores of 70.0, 72.5, 74.1, and 77.9 on \textit{Dynamic Media}, \textit{Gaming}, \textit{Music}, and \textit{Structured Speech}, respectively. The improvement is primarily recall-driven, with recall rising to 88.5, 83.8, 90.1, and 90.3, while precision remains comparatively lower. On \textit{Music}, for example, the model achieves 90.1 recall but 62.9 precision. This indicates that imitating the \textit{AudioChaps-CoT} targets establishes a strong high-recall prior, providing an effective starting point for subsequent GRPO calibration.

\paragraph{GRPO calibration delivers the best overall trade-off.}
Comparing AudioChaps-SFT-8B with AudioChaps-R1-8B isolates the role of GRPO once a structured reasoning prior has been established through the CoT cold start. Rather than simply increasing boundary sensitivity, GRPO corrects the permissive operating point induced by SFT: precision rises by 12.1 points on \textit{Dynamic Media}, 11.2 on \textit{Gaming}, 20.7 on \textit{Music}, and 2.5 on \textit{Structured Speech}, while recall decreases to varying degrees. The benefit is largest where SFT leaves the most precision headroom. On \textit{Music}, GRPO trades 4.5 points of recall for a 20.7-point precision gain, increasing F1 by 10.5 points. On \textit{Structured Speech}, SFT is already well calibrated, so the 2.5-point precision gain is offset by a 4.2-point recall reduction and F1 remains effectively unchanged. Overall, AudioChaps-R1-8B achieves the best F1 on three of the four subtypes and remains effectively tied with AudioChaps-SFT-8B on \textit{Structured Speech}. These results clarify how GRPO operates once an evidence-grounded reasoning initialization and a consistent output format have been established. SFT teaches the model to interpret the audio in relation to the chapter-boundary question and to produce well-structured reasoning, while GRPO improves precision by calibrating the resulting decisions toward creator-authored annotations, primarily by suppressing unsupported boundary calls. A similar precision-oriented shift is observed with the MOSS-Think backbone, as discussed below. Full results are provided in Supplementary Table~\ref{tab_detailed_ablation}.

\paragraph{Backbone generalization of AudioChaps.}
AudioChaps is designed not as a replacement for a particular base LALM, but as a backbone-agnostic GRPO-based alignment framework that strengthens the chapterization ability already present in that model. To evaluate whether its gains extend beyond AF3-Think-8B, we apply AudioChaps to MOSS-Audio-8B-Thinking (MOSS-Think-8B)~\citep{yang2026mossaudiotechnicalreport}, which exhibits strong zero-shot chapterization performance. Because MOSS-Think-8B has already undergone supervised instruction tuning, an explicit reasoning cold start, and reinforcement learning for audio reasoning, it already provides a structured reasoning prior and consistent reasoning format. We therefore apply the AudioChaps GRPO alignment stage directly. AudioChaps further improves MOSS-Think-8B across all four acoustic subtypes, increasing F1 from 69.3 to 79.2 on \textit{Dynamic Media}, 72.9 to 78.2 on \textit{Gaming}, 75.7 to 85.6 on \textit{Music}, and 76.5 to 82.8 on \textit{Structured Speech}. The gain is accompanied by the same precision-oriented shift observed in the AF3-Think-8B stage-wise ablation: average precision rises from 63.3 to 82.4, while recall decreases from 88.0 to 80.6, increasing average F1 from 73.6 to 81.4. These results show that, when a suitable reasoning prior and output format are already present, task-specific GRPO can further improve the native chapterization ability of a capable LALM by suppressing unsupported boundary calls and calibrating predictions toward creator-authored annotations. Full results are provided in Table~\ref{tab_moss_backbone_generalization} and Section~\ref{apx_backbone_generalization} of the supplementary material.

\paragraph{Comparison with additional multimodal and omni LLMs.}
We further compare AudioChaps against two strong zero-shot baselines on AudioChaps-Eval: Qwen3-Omni-30B-A3B-Thinking~\citep{xu2025qwen3omnitechnicalreport} and Gemini 2.5 Flash~\citep{comanici2025gemini25pushingfrontier}. MOSS-Think-AudioChaps-R1-8B outperforms both larger general-purpose models, achieving an average F1 of 81.44, compared with 76.75 for Gemini 2.5 Flash and 75.29 for Qwen3-Omni. This corresponds to gains of 4.69 and 6.15 F1 points, respectively. These results show that AudioChaps enables an 8B-parameter LALM to outperform larger zero-shot models by aligning its chapterization decisions with creator-authored annotations.

\section{Conclusion}
We study audio chapterization as a concrete setting in which LALMs can move beyond benchmark performance toward practical media workflows. By treating chapter-boundary placement as editorial judgment rather than a fixed acoustic event, we align end-to-end LALMs with creator-authored chapter annotations using three resources: AudioChaps-Alignment, AudioChaps-CoT, and AudioChaps-Eval. Our two-stage training recipe combines a CoT SFT cold start, which establishes a structured high-recall reasoning prior, with GRPO, which improves precision by suppressing unsupported boundary calls and calibrating predictions toward creator-authored annotations. AudioChaps-R1 substantially improves over its AF3-Think-8B backbone and surpasses the larger Step-Audio-R1 model used to produce its acoustic perception supervision, despite using roughly one quarter of the parameters. 
\paragraph{Limitations.}
The current framework does not natively support long-context audio modelling. This limits its ability to generate globally informed chapter titles and segment summaries. Extending AudioChaps to native long-context modelling and full chapter generation is therefore an important direction toward deployment in production media workflows.

\section*{Acknowledgments}

This research was supported by the EPSRC and BBC Prosperity Partnership
``AI4ME: Future Personalized Object Based Media Experiences Delivered
at Scale Anywhere'' (EP/V038087/1). The authors also acknowledge the
computational resources provided by Isambard-AI, part of the National
AI Research Resource (AIRR)~\citep{mcintoshsmith2024isambardaileadershipclasssupercomputer}.
Isambard-AI is operated by the University of Bristol and funded by the
UK Government's Department for Science, Innovation and Technology
(DSIT) through UK Research and Innovation (UKRI) and the Science and
Technology Facilities Council (STFC), under grant ST/AIRR/I-A-I/1023.


\bibliography{custom}

\newpage

\appendix
\section{Training Details}
\subsection{Training Prompt Template}
\label{apx_train_prompt}
Across both training stages, we adopt a unified prompt template that frames chapterization as a single-choice question. Each prompt presents the binary query introduced in Section 3.1, augmented with two prescriptive elements: (i) an instruction to engage in deliberate, human-like internal reflection, using natural cognitive markers such as ``let me think,'' ``wait,'' or ``let's break it down,'' and to perform self-verification before committing to a decision; and (ii) a strict output specification requiring the reasoning trace to be enclosed within \texttt{<think>...</think>} tags and the final answer, restricted to a single option letter, within \texttt{<answer>...</answer>} tags. The full prompt is shown in Figure~\ref{fig_prompt}.

\begin{figure*}[t]
\centering
\begin{tcolorbox}[
    width=\textwidth,
    colback=gray!5,
    colframe=gray!50!black,
    title=Training prompt template,
    fonttitle=\bfseries,
    boxrule=0.5pt,
    arc=2pt,
]
\small\ttfamily
Is there a chapter boundary present in this audio? Choose one from the following choices:\\
(A) Yes\\
(B) No\\[4pt]
Please think about this question as if you were a human pondering deeply. Engage in an internal dialogue using expressions such as `let me think', `wait', `Hmm', `oh, I see', `let's break it down', etc, or other natural language thought expressions. It's encouraged to include self-reflection or verification in the reasoning process. Provide your detailed reasoning between the <think> </think> tags, and then give your final answer between the <answer> </answer> tags. Please provide only the single option letter (e.g., A, B) within the <answer> </answer> tags.
\end{tcolorbox}
\caption{The unified prompt template used in both the SFT cold-start and GRPO stages. The model is asked a binary single-choice question, instructed to reason with natural reflective markers inside \texttt{<think>} tags, and to emit a single option letter inside \texttt{<answer>} tags.}
\label{fig_prompt}
\end{figure*}

\subsection{Hyperparameters and Training Budget}
\label{apx_hyperparams_budget}
 
Table~\ref{tab_hyperparams} lists the hyperparameters for both training stages. In terms of compute, the SFT stage was run on a single node with 8 NVIDIA H200-140GB GPUs and completed in approximately 4 hours. GRPO alignment was run on 8 nodes, each with 4 NVIDIA GH200-96GB, and completed in approximately 10 hours. Both stages fine-tune the 8B-parameter backbone rather than training from scratch, keeping the overall budget modest relative to pretraining a model of comparable capability.
 
\begin{table}[t]
\centering
\small
\begin{tabular}{lcc}
\toprule
\textbf{Hyperparameter} & \textbf{SFT} & \textbf{GRPO} \\
\midrule
Epochs                      & 2        & 1        \\
Learning rate               & $1\mathrm{e}{-6}$ & $1\mathrm{e}{-6}$ \\
Per-device batch size       & 1        & 1        \\
Gradient accumulation       & 4        & 1        \\
Max completion length       & ---      & 768      \\
Generations per prompt ($G$) & ---     & 8        \\
KL coefficient ($\beta$)    & ---      & 0.04     \\
Max gradient norm           & 5        & 5        \\
Precision                   & bf16     & bf16     \\
Attention                   & FA-2     & FA-2     \\
DeepSpeed                   & ZeRO-2   & ZeRO-2   \\
\bottomrule
\end{tabular}
\caption{Training hyperparameters for the SFT cold-start and GRPO alignment stages.}
\label{tab_hyperparams}
\end{table}

\section{Detailed Ablation Results}
\label{apx_detailed_ablation}

This section provides the full per-subtype, per-metric breakdown behind the F1 summary reported in the main paper Table 3. Table~\ref{tab_detailed_ablation} lists accuracy, precision, recall, and F1 for the AF3-Think-8B base and the three variants of our pipeline (R1-Zero, SFT, and the full R1 model) across all four content subcategories, allowing the precision and recall behaviour discussed in Section 4.3 to be read off directly.

\begin{table}[t]
\centering
\small
\setlength{\tabcolsep}{3.5pt}
\begin{tabular}{p{18.0ex}lcccc}
\toprule
{\mbox{ }\hspace{-1.5ex}\textbf{Model}} &
\textbf{sc} & \textbf{Acc} & \textbf{Pre} &
\textbf{Rec} & \textbf{F1} \\
\midrule

& DM & 55.8 & 47.5 & 23.7 & 31.6 \\
{\mbox{ }\hspace{-1.5ex}AF3-Think-8B}
& G & 50.1 & 45.9 & 54.6 & 49.9 \\
{\mbox{ }\hspace{-1.5ex}(Base-LALM)}
& M & 62.3 & 64.0 & 3.2 & 6.0 \\
& SS & 44.7 & 46.4 & 19.1 & 27.0 \\
\cline{2-6}
& \textit{Avg} & \textit{53.2} & \textit{51.0} &
\textit{25.2} & \textit{28.6} \\
\midrule

& DM & 61.6 & 55.0 & 61.0 & 57.8 \\
{\mbox{ }\hspace{-1.5ex}AudioChaps-R1-}
& G & 65.3 & 60.3 & 69.2 & 64.4 \\
{\mbox{ }\hspace{-1.5ex}Zero-8B}
& M & 75.6 & 82.1 & 46.1 & 59.1 \\
& SS & 60.0 & 62.8 & 62.9 & 62.9 \\
\cline{2-6}
& \textit{Avg} & \textit{65.6} & \textit{65.1} &
\textit{59.8} & \textit{61.1} \\
\midrule

& DM & 67.3 & 57.9 & \textbf{88.5} & 70.0 \\
{\mbox{ }\hspace{-1.5ex}AudioChaps-SFT-}
& G & 71.1 & 63.9 & \textbf{83.8} & 72.5 \\
{\mbox{ }\hspace{-1.5ex}8B}
& M & 75.9 & 62.9 & \textbf{90.1} & 74.1 \\
& SS & 72.5 & 68.5 & \textbf{90.3} & \textbf{77.9} \\
\cline{2-6}
& \textit{Avg} & \textit{71.7} & \textit{63.3} &
\textbf{\textit{88.2}} & \textit{73.6} \\
\midrule

& DM & \textbf{75.9} & \textbf{70.0} & 77.2 &
\textbf{73.4} \\
{\mbox{ }\hspace{-1.5ex}AudioChaps-R1-}
& G & \textbf{77.6} & \textbf{75.1} & 75.9 &
\textbf{75.5} \\
{\mbox{ }\hspace{-1.5ex}8B}
& M & \textbf{88.1} & \textbf{83.6} & 85.6 &
\textbf{84.6} \\
& SS & \textbf{73.6} & \textbf{71.0} & 86.0 & 77.8 \\
\cline{2-6}
& \textit{Avg} & \textbf{\textit{78.8}} &
\textbf{\textit{74.9}} & \textit{81.2} &
\textbf{\textit{77.8}} \\
\bottomrule
\end{tabular}
\caption{
Detailed ablation results corresponding to Table~\ref{tab:ablation_f1} in the main paper across four audio chapterization subcategories (sc): \textit{Dynamic Media} (DM), \textit{Gaming} (G), \textit{Music} (M), and \textit{Structured Speech} (SS). AF3-Think-8B is the unaligned base LALM, which lacks a sufficiently strong reasoning prior and does not reliably follow the target structured output format. AudioChaps-R1-Zero-8B applies GRPO directly to AF3-Think-8B without an SFT cold start. AudioChaps-SFT-8B uses supervised fine-tuning to establish the target reasoning behavior and structured output format, while AudioChaps-R1-8B subsequently applies GRPO to calibrate the model's chapter-boundary decisions. This ablation isolates the contribution of the SFT cold start and the subsequent GRPO alignment stage. The macro average (Avg) is reported over the four subcategories.
}
\label{tab_detailed_ablation}
\end{table}


\section{Backbone Generalization of AudioChaps}
\label{apx_backbone_generalization}

Our primary contribution is not a specific backbone, but a GRPO-based framework for adapting LALMs to creator-authored editorial annotations for audio chapterization. We instantiate AudioChaps on both AF3-Think-8B~\citep{goel2025audio} and MOSS-Audio-8B-Thinking (MOSS-Think-8B)~\citep{yang2026mossaudiotechnicalreport}, adapting the initialization stage to the capabilities of each model. AF3-Think-8B requires a CoT SFT cold start to establish the target reasoning format before GRPO. In contrast, MOSS-Think-8B has already undergone an explicit reasoning cold start and reinforcement learning for audio reasoning. It therefore provides a strong reasoning prior and produces outputs in the required structured format, allowing us to apply the AudioChaps GRPO stage directly.

A consistent precision-recall trend emerges across the two backbones.
As shown in Table~\ref{tab_detailed_ablation}, the AF3-based SFT model
is strongly recall-oriented, achieving 88.2 macro-average recall but
only 63.3 precision. Subsequent GRPO calibration raises precision to
74.9 and F1 from 73.6 to 77.8, while reducing recall to 81.2. The same behaviour is observed with MOSS-Think-8B
(Table~\ref{tab_moss_backbone_generalization}): the base LALM
(MOSS-Think-8B) achieves 88.0 recall and 63.3 precision, whereas the
AudioChaps-adapted model (MOSS-Think-AudioChaps-R1-8B) raises precision
to 82.4 and F1 from 73.6 to 81.4, with recall decreasing to 80.6.

The MOSS-based AudioChaps model(MOSS-Think-AudioChaps-R1-8B) improves F1 across all four
subcategories, with gains of 9.9 points on \textit{Dynamic Media},
5.3 on \textit{Gaming}, 9.9 on \textit{Music}, and 6.3 on
\textit{Structured Speech}. These results indicate that GRPO plays a
similar calibration role across distinct LALM families: it corrects an
overly recall-heavy decision policy and produces a more selective and
balanced chapter-boundary detector. At the same time, these results show that AudioChaps can further improve the native chapterization ability of different LALMs, regardless of their initial capabilities, by calibrating their boundary decisions toward creator-authored annotations.

\begin{table}[t]
\centering
\small
\setlength{\tabcolsep}{3.5pt}
\begin{tabular}{p{18.0ex}lcccc}
\toprule
{\mbox{ }\hspace{-1.5ex}\textbf{Model}} &
\textbf{sc} & \textbf{Acc} & \textbf{Pre} & \textbf{Rec} & \textbf{F1} \\
\midrule

\rowcolor{gray!12}
& DM & 55.8 & 47.5 & 23.7 & 31.6 \\
\rowcolor{gray!12}
{\mbox{ }\hspace{-1.5ex}AF3-Think-8B}
& G & 50.1 & 45.9 & 54.6 & 49.9 \\
\rowcolor{gray!12}
{\mbox{ }\hspace{-1.5ex}(Base-LALM)}
& M & 62.3 & 64.0 & 3.2 & 6.0 \\
\rowcolor{gray!12}
& SS & 44.7 & 46.4 & 19.1 & 27.0 \\
\cmidrule(l){2-6}
\rowcolor{gray!12}
& \textit{Avg} & 53.2 & 51.0 & 25.2 & 28.6 \\

\midrule

\rowcolor{gray!12}
& DM & 75.9 & 70.0 & 77.2 & 73.4 \\
\rowcolor{gray!12}
{\mbox{ }\hspace{-1.5ex}AudioChaps-R1-8B}
& G & 77.6 & 75.1 & 75.9 & 75.5 \\
\rowcolor{gray!12}
& M & 88.1 & 83.6 & 85.6 & 84.6 \\
\rowcolor{gray!12}
& SS & 73.6 & 71.0 & 86.0 & 77.8 \\
\cmidrule(l){2-6}
\rowcolor{gray!12}
& \textit{Avg} & 78.8 & 74.9 & 81.2 & 77.8 \\

\midrule
\midrule

& DM & 67.0 & 57.8 & \textbf{86.4} & 69.3 \\
{\mbox{ }\hspace{-1.5ex}MOSS-Think-8B}
& G & 70.9 & 63.3 & \textbf{85.9} & 72.9 \\
{\mbox{ }\hspace{-1.5ex}(Base-LALM)}
& M & 77.9 & 65.3 & \textbf{89.9} & 75.7 \\
& SS & 70.4 & 66.7 & \textbf{89.6} & 76.5 \\
\cmidrule(l){2-6}
& \textit{Avg} & 71.5 & 63.3 & \textbf{88.0} & 73.6 \\

\midrule

& DM & \textbf{82.4} & \textbf{80.6} & 77.9 & \textbf{79.2} \\
{\mbox{ }\hspace{-1.5ex}MOSS-Think-}
& G & \textbf{80.8} & \textbf{80.6} & 75.9 & \textbf{78.2} \\
{\mbox{ }\hspace{-1.5ex}AudioChaps-R1-8B}
& M & \textbf{89.0} & \textbf{85.6} & 85.6 & \textbf{85.6} \\
& SS & \textbf{81.5} & \textbf{82.7} & 82.9 & \textbf{82.8} \\
\cmidrule(l){2-6}
& \textit{Avg} & \textbf{83.4} & \textbf{82.4} & 80.6 & \textbf{81.4} \\

\bottomrule
\end{tabular}
\caption{
Backbone-generalization results across audio chapterization subcategories
(sc): \textit{Dynamic Media} (DM), \textit{Gaming} (G), \textit{Music}
(M), and \textit{Structured Speech} (SS). Grey rows report the AF3-based
models for reference. Across both AF3 and MOSS, GRPO alignment improves
precision (Pre), accuracy (Acc), and F1 while reducing the high-recall
behaviour of the preceding model. \textit{Avg} denotes the macro-average
over subcategories.
}
\label{tab_moss_backbone_generalization}
\end{table}

\section{Comparison with Larger General-Purpose Multimodal and Omni-Modal Models}
\label{apx_comparision_large_mllm}
We further compare AudioChaps with two more recent and substantially larger
audio-language models, Gemini~2.5~Flash~\citep{comanici2025gemini25pushingfrontier} and Qwen3-Omni-30B-A3B-Thinking~\citep{xu2025qwen3omnitechnicalreport}.
Table~\ref{tab_advanced_lalm_comparison} shows that the 8B-parameter
AudioChaps models remain competitive despite their smaller scale.
AudioChaps-R1 with the MOSS backbone achieves the best macro-average
accuracy, precision, and F1, reaching 83.4, 82.4, and 81.4, respectively.
Compared with Gemini~2.5~Flash, it improves macro-average F1 by 4.6 points,
and compared with Qwen3-Omni-30B-A3B-Thinking, it improves F1 by 6.1 points.
Gemini obtains the highest macro-average recall at 82.7, indicating a more
recall-oriented prediction policy, whereas MOSS-Think-AudioChaps-R1-8B provides a better
precision--recall balance. The advantage of MOSS-Think-AudioChaps-R1-8B is consistent
across subcategories, achieving the highest F1 on Dynamic Media, Gaming,
Music, and Structured Speech. These results suggest that task-specific alignment can enable a compact open model to outperform substantially larger general-purpose LALMs on audio chapterization. This comparison evaluates the benefit of specialized alignment rather than the ultimate capability of stronger or larger backbones, which may improve further when adapted with AudioChaps.

\begin{table}[t]
\centering
\small
\setlength{\tabcolsep}{3.5pt}
\begin{tabular}{p{18.0ex}lcccc}
\toprule
{\mbox{ }\hspace{-1.5ex}\textbf{Model}} &
\textbf{sc} & \textbf{Acc} & \textbf{Pre} &
\textbf{Rec} & \textbf{F1} \\
\midrule

& DM & 72.5 & 66.7 & \textbf{78.8} & 72.2 \\
{\mbox{ }\hspace{-1.5ex}Gemini-2.5-Flash}
& G  & 73.0 & 67.9 & \textbf{82.5} & 74.5 \\
{\mbox{ }\hspace{-1.5ex}(Zero-Shot)}
& M  & 81.0 & 74.9 & 81.4 & 78.0 \\
& SS & 78.3 & 77.3 & \textbf{88.0} & 82.3 \\
\cline{2-6}
& \textit{Avg} & \textit{76.2} & \textit{71.7} &
\textbf{\textit{82.7}} & \textit{76.8} \\

\midrule

& DM & 73.1 & 68.6 & 69.2 & 68.9 \\
{\mbox{ }\hspace{-1.5ex}Qwen3-Omni-30B-}
& G  & 75.6 & 74.9 & 69.7 & 72.2 \\
{\mbox{ }\hspace{-1.5ex}A3B-Thinking}
& M  & 85.3 & 80.6 & 81.0 & 80.8 \\
{\mbox{ }\hspace{-1.5ex}(Zero-Shot)}
& SS & 77.6 & 78.7 & 79.8 & 79.2 \\
\cline{2-6}
& \textit{Avg} & \textit{77.9} & \textit{75.7} &
\textit{74.9} & \textit{75.3} \\
\midrule

& DM & 67.0 & 57.8 & \textbf{86.4} & 69.3 \\
{\mbox{ }\hspace{-1.5ex}MOSS-Think-8B}
& G & 70.9 & 63.3 & \textbf{85.9} & 72.9 \\
{\mbox{ }\hspace{-1.5ex}(Zero-Shot)}
& M & 77.9 & 65.3 & \textbf{89.9} & 75.7 \\
& SS & 70.4 & 66.7 & \textbf{89.6} & 76.5 \\
\cmidrule(l){2-6}
& \textit{Avg} & 71.5 & 63.3 & \textbf{88.0} & 73.6 \\

\midrule
& DM      & 60.2 & 55.3 & 62.7 & 58.8 \\
{\mbox{ }\hspace{-1.5ex}Step-Audio-R1-32B\hspace{-2ex}\mbox{ }}
& G             & 60.1 & 59.1 & 53.4 & 56.1 \\
{\mbox{ }\hspace{-1.5ex}(Zero-Shot)}
& M              & 62.4 & 53.1 & 51.9 & 52.5 \\
& SS  & 65.6 & 70.1 & 69.7 & 69.9 \\
\cmidrule(l){2-6}
&\textit{Avg}& 62.1 & 59.4 & 59.4 & 59.3 \\
\midrule
& DM       & 55.8 & 47.5 & 23.7 & 31.6 \\
{\mbox{ }\hspace{-1.5ex}AF3-Think-8B}
& G             & 50.1 & 45.9 & 54.6 & 49.9 \\
{\mbox{ }\hspace{-1.5ex}(Zero-Shot)}
& M              & 62.3 & 64.0 &  3.2 &  6.0 \\
& SS  & 44.7 & 46.4 & 19.1 & 27.0 \\
\cmidrule(l){2-6}
&\textit{Avg}& 53.2 & 51.0 & 25.2 & 28.6 \\

\midrule

& DM & 75.9 & 70.0 & 77.2 & 73.4 \\
{\mbox{ }\hspace{-1.5ex}AudioChaps-R1-8B}
& G & 77.6 & 75.1 & 75.9 & 75.5 \\
{\mbox{ }\hspace{-1.5ex}}
& M & 88.1 & 83.6 & 85.6 & 84.6 \\
& SS & 73.6 & 71.0 & 86.0 & 77.8 \\
\cline{2-6}
& \textit{Avg} & \textit{78.8} & \textit{74.9} &
\textit{81.2} & \textit{77.8} \\

\midrule

& DM & \textbf{82.4} & \textbf{80.6} & 77.9 &
\textbf{79.2} \\
{\mbox{ }\hspace{-1.5ex}MOSS-Think-}
& G & \textbf{80.8} & \textbf{80.6} & 75.9 &
\textbf{78.2} \\
{\mbox{ }\hspace{-1.5ex}AudioChaps-R1-8B}
& M & \textbf{89.0} & \textbf{85.6} & \textbf{85.6} &
\textbf{85.6} \\
& SS & \textbf{81.5} & \textbf{82.7} & 82.9 &
\textbf{82.8} \\
\cline{2-6}
& \textit{Avg} & \textbf{\textit{83.4}} &
\textbf{\textit{82.4}} & \textit{80.6} &
\textbf{\textit{81.4}} \\

\bottomrule
\end{tabular}
\caption{
Comparison with additional multimodal and omni-modal LLMs on
AudioChaps-Eval across four audio chapterization subcategories:
\textit{Dynamic Media} (DM), \textit{Gaming} (G), \textit{Music} (M),
and \textit{Structured Speech} (SS). Accuracy (Acc), precision (Pre),
recall (Rec), and F1 are reported as percentages, together with the
macro-average (Avg) over the four subcategories. Gemini-2.5-Flash,
Qwen3-Omni-30B-A3B-Thinking, MOSS-Think-8B, Step-Audio-R1-32B, and
AF3-Think-8B are evaluated in the zero-shot setting. The two
AudioChaps-R1 variants apply our alignment framework to the AF3-Think
and MOSS-Think backbones, respectively.
}
\label{tab_advanced_lalm_comparison}
\end{table}

\section{Statistical Significance of the Main Improvement}
\label{apx_bootstrap}

We assess the uncertainty of the performance difference between AF3-Think-8B and AudioChaps-R1-8B using a paired non-parametric bootstrap. Because multiple evaluation clips may originate from the same source video and are therefore not independent, we resample source videos rather than individual clips. In each of 10,000 bootstrap iterations, the same sampled videos are used to evaluate both models, preserving the paired comparison. For each resample, we compute positive-class F1 independently for each of the four acoustic subcategories and report their macro-average, matching the aggregation used in the main results.

AudioChaps-R1-8B obtains a macro-F1 of 77.8, compared with 28.6 for AF3-Think-8B. The resulting improvement is 49.2 F1 points, with a 95\% bootstrap confidence interval of [46.2, 52.2] and a two-sided bootstrap $p$-value below $10^{-4}$. The confidence interval excludes zero by a wide margin. Improvements are also significant within every acoustic subtype: Dynamic Media, $+41.8$ [37.5, 46.3]; Gaming, $+25.6$ [19.9, 31.7]; Music, $+78.6$ [73.9, 83.0]; and Structured Speech, $+50.8$ [42.4, 58.8]. The analysis includes 15,963 paired clips from 749 source videos.

\section{Full-Length Audio Chapterization Evaluation}
\label{apx_continuous_eval}

The clip-level AudioChaps-Eval benchmark provides a controlled setting for isolating chapter-boundary judgment under balanced positive and negative sampling. We additionally evaluate whether these local decisions can be converted into recording-level chapter boundaries on uncropped, full-length audio. This setting is more challenging because chapter boundaries are sparse, each recording may contain multiple boundaries, and overlapping window-level predictions must be consolidated into a sequence of recording-level estimates.

\subsection{Evaluation Set and Sliding-Window Inference}

We evaluate all models on the common intersection of 40 full-length recordings, with 10 recordings from each of the four acoustic subcategories. The recordings have a mean duration of 49 minutes 29 seconds, yielding approximately 33 hours of long-form audio in total, and together contain 387 reference chapter boundaries.

Each recording is divided into overlapping 60-second windows with a 20-second hop. A final tail window is added when necessary to ensure complete coverage of the recording. This produces 5,875 evaluation windows across the 40 recordings.

\subsection{Recording-Level Decoding}

For models that output only a binary boundary verdict, we use a
\emph{window-run decoder}. Consecutive windows predicted as positive are
grouped into a contiguous run, and one boundary is placed at the midpoint
of the run's temporal coverage. Estimated boundaries separated by fewer
than 20 seconds are subsequently merged by taking their median. This
procedure prevents a single transition observed in several overlapping
windows from being counted multiple times.

\subsection{Matching and Metrics}

Predicted and reference boundaries are matched one-to-one under a
$\pm 10$-second tolerance. This tolerance follows from the training-data
construction: positive 60-second crops place the reference boundary
within the central 20 seconds, so a window-centre estimate localizes the
boundary to within at most 10 seconds. Candidate matches within this
tolerance are ordered by absolute temporal distance and greedily assigned
without reusing either a prediction or a reference boundary.

We compute precision, recall, and F1 independently for each recording,
macro-average them within each acoustic subcategory, and then average
equally across the four subcategories. We additionally report
\mbox{dev-R2E}, the pooled median distance from each reference boundary
to its nearest predicted boundary, and the ratio between the numbers of
predicted and reference boundaries. As a content-independent baseline,
we place one boundary every 180 seconds.

\begin{table}[t]
\centering
\small
\setlength{\tabcolsep}{3.5pt}
\begin{tabular}{p{18.0ex}lcccc}
\toprule
{\mbox{ }\hspace{-1.5ex}\textbf{Model}} &
\textbf{sc} & \textbf{Pre} & \textbf{Rec} &
\textbf{F1} & \textbf{dev-R2E$\downarrow$} \\
\midrule

& DM & 18.9 & 13.5 & 10.7 & -- \\
{\mbox{ }\hspace{-1.5ex}AF3-Think-8B}
& G  & 3.4 & 10.5 & 4.6 & -- \\
{\mbox{ }\hspace{-1.5ex}(Window)}
& M  & 0.4 & 1.7 & 0.7 & -- \\
& SS & 11.8 & 12.7 & 10.0 & -- \\
\cline{2-6}
& \textit{Overall} & \textit{8.6} & \textit{9.6} &
\textit{6.5} & \textit{38.0 s} \\

\midrule

& DM & 32.6 & 38.6 & 34.5 & -- \\
{\mbox{ }\hspace{-1.5ex}AudioChaps-R1-8B}
& G  & 20.0 & 38.3 & 24.6 & -- \\
{\mbox{ }\hspace{-1.5ex}(Window)}
& M  & 58.1 & 73.7 & 63.7 & -- \\
& SS & 25.8 & 32.3 & 27.5 & -- \\
\cline{2-6}
& \textit{Overall} & \textit{34.1} & \textit{45.7} &
\textit{37.6} & \textit{10.0 s} \\

\midrule

& DM & 17.4 & 26.7 & 20.4 & -- \\
{\mbox{ }\hspace{-1.5ex}MOSS-Think-8B}
& G  & 6.5 & 19.9 & 8.1 & -- \\
{\mbox{ }\hspace{-1.5ex}(Window)}
& M  & 35.8 & 63.4 & 44.2 & -- \\
& SS & 30.1 & 38.5 & 32.8 & -- \\
\cline{2-6}
& \textit{Overall} & \textit{22.4} & \textit{37.1} &
\textit{26.4} & \textit{15.0 s} \\

\midrule

& DM & 53.3 & 41.1 & 42.4 & -- \\
{\mbox{ }\hspace{-1.5ex}MOSS-Think-}
& G  & 26.2 & 27.4 & 23.2 & -- \\
{\mbox{ }\hspace{-1.5ex}AudioChaps-R1-8B}
& M  & 60.4 & 73.0 & 65.5 & -- \\
{\mbox{ }\hspace{-1.5ex}(Window)}
& SS & 52.6 & 55.5 & 53.7 & -- \\
\cline{2-6}
& \textit{Overall} & \textit{48.2} & \textit{49.3} &
\textit{46.2} & \textit{8.0 s} \\

\midrule

& DM & 12.0 & 13.8 & 12.5 & -- \\
{\mbox{ }\hspace{-1.5ex}Fixed Interval}
& G  & 8.8 & 8.6 & 8.2 & -- \\
{\mbox{ }\hspace{-1.5ex}(180 s)}
& M  & 9.4 & 15.1 & 11.0 & -- \\
& SS & 5.9 & 8.7 & 6.2 & -- \\
\cline{2-6}
& \textit{Overall} & \textit{9.0} & \textit{11.6} &
\textit{9.5} & \textit{49.0 s} \\

\bottomrule
\end{tabular}
\caption{
Full-length audio chapter-boundary detection across four acoustic subcategories: \textit{Dynamic Media} (DM), \textit{Gaming} (G), \textit{Music} (M), and \textit{Structured Speech} (SS). Predictions are matched one-to-one with reference boundaries under a $\pm 10$-second tolerance. Precision (Pre), recall (Rec), and F1 are computed per recording, macro-averaged within each subcategory, and then averaged equally across the four subcategories. \mbox{dev-R2E} denotes the pooled median distance from each reference boundary to its nearest prediction and is reported only in the \textit{Overall} row; lower values indicate better localization. The window-run decoder merges contiguous positive windows into a single recording-level boundary estimate.
}
\label{tab_long_audio_full}
\end{table}

\subsection{Results}

The fixed-interval baseline reaches only 9.5 average F1, while
AF3-Think-8B obtains 6.5 F1 and predicts no boundaries on 15 of the
40 recordings. AudioChaps-R1-8B improves average F1 to 37.6 and reduces
the pooled median reference-to-estimate deviation from 38 to 10 seconds,
showing that clip-level training transfers to full-length recordings.

With the window-run decoder, MOSS-Think-AudioChaps-R1-8B improves over
the base LALM, MOSS-Think-8B, from 26.4 to 46.2 F1, primarily through
a precision gain from 22.4 to 48.2. Overall, AudioChaps improves
full-length chapterization across both backbones, demonstrating that
locally trained boundary judgments can be consolidated into effective
recording-level chapter predictions.

\section{Training and Evaluation Dataset Statistics}
\label{apx_data_stats}
\begin{figure*}[t]
    \centering
    \includegraphics[width=\textwidth]{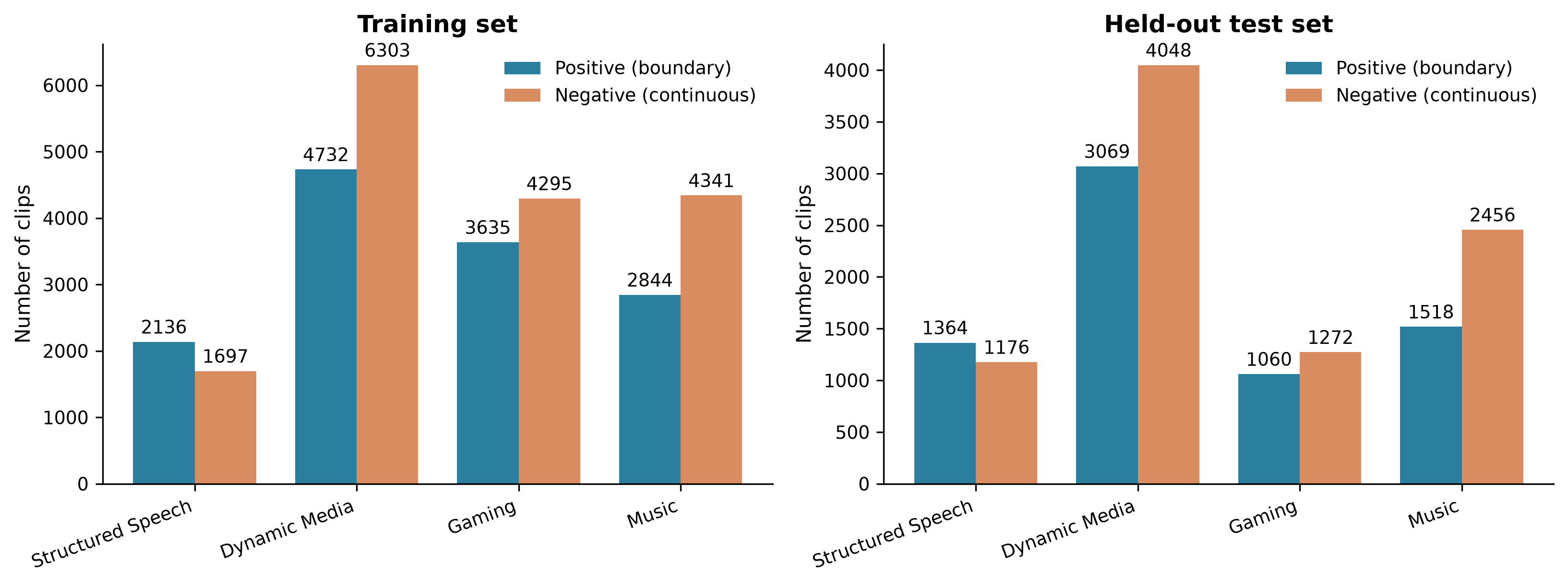}
    \caption{Per-subtype clip counts for the training (AudioChaps-Alignment) and held-out test (AudioChaps-Eval) splits, split by positive (boundary-spanning) and negative (within-chapter) clips. Both splits follow the same four-way stratification, with no source-video overlap between the training set and the held-out test set.}
    \label{fig_data_stats}
\end{figure*}

As discussed in Section 4.1, our data is curated from VidChapters-7M~\citep{yang2023vidchapters7mvideochaptersscale} and stratified into four acoustic subtypes. Here we provide a finer breakdown of the dataset by subcategory and by positive/negative clip balance. Figure~\ref{fig_data_stats} reports the per-subtype clip counts for the training and held-out test splits, together with the positive-to-negative ratio within each subtype.

\section{Sample Output Comparison: AF3-Think vs. AudioChaps-R1 with Human Evaluation}
\label{apx_human_eval}
 
\begin{figure*}[t]
    \centering
    \includegraphics[width=\textwidth]{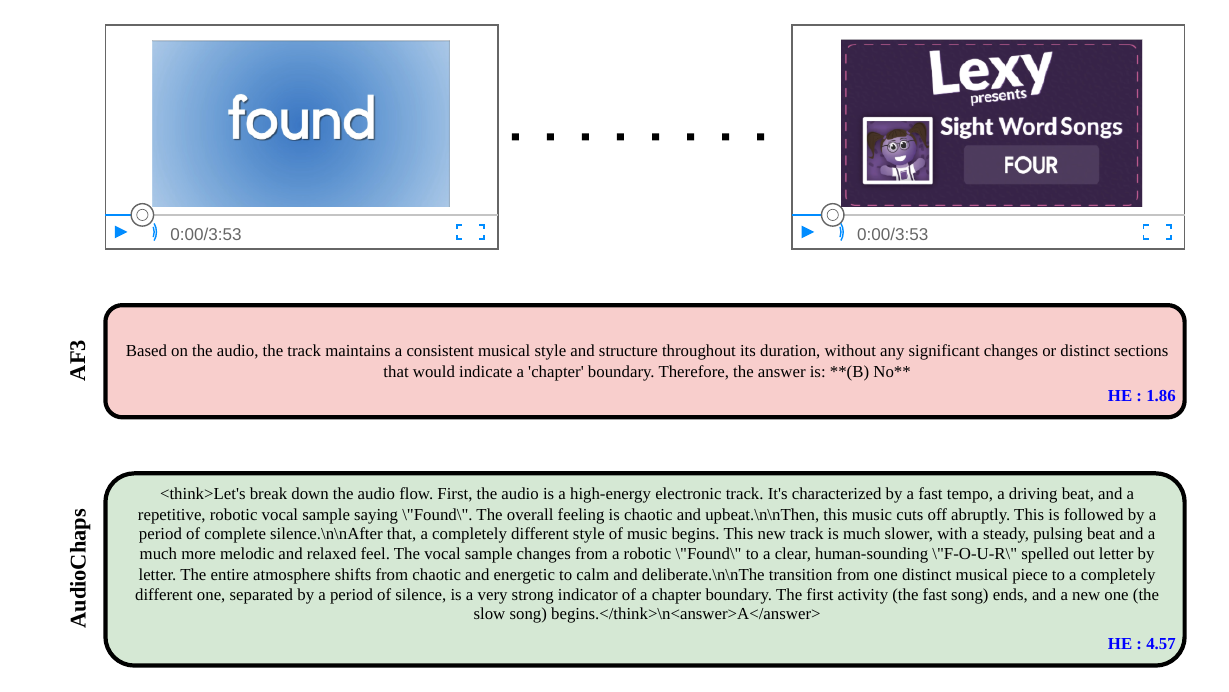}
    \caption{\small Qualitative comparison of AF3-Think-8B and AudioChaps-R1-8B on human-evaluation (HE).}    
    \label{fig_rating1}
\end{figure*}

\begin{figure*}[t]
    \centering
    \includegraphics[width=\textwidth]{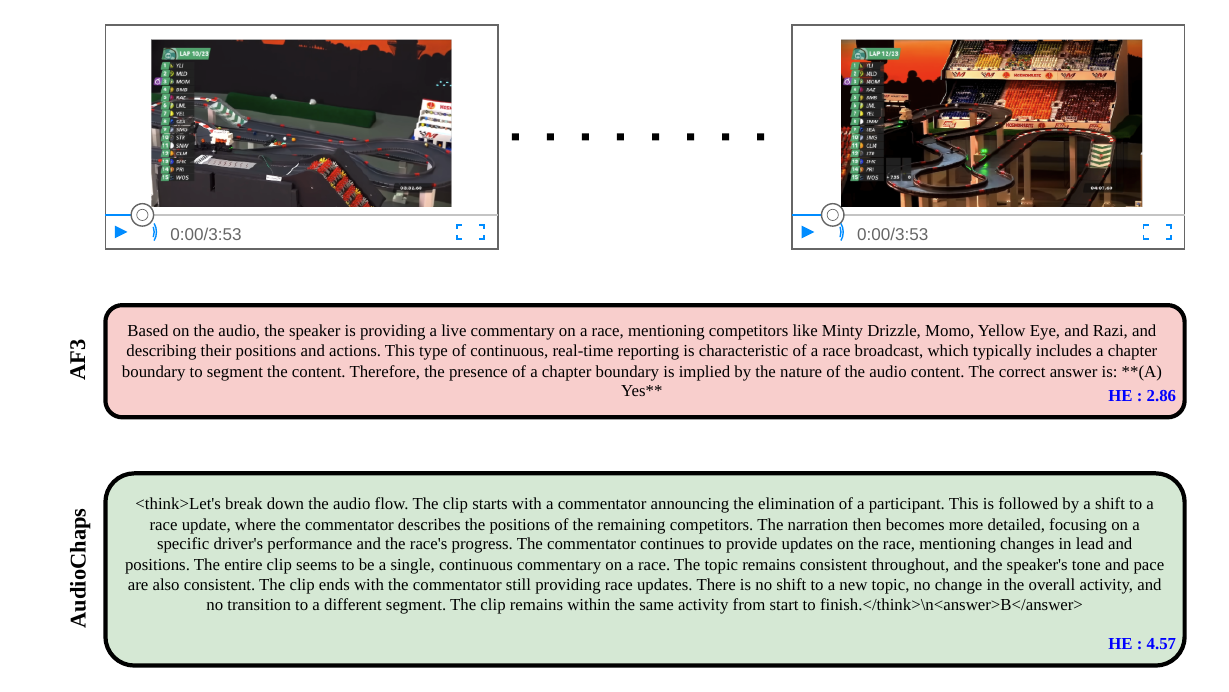}
    \caption{\small Qualitative comparison of AF3-Think-8B and AudioChaps-R1-8B on human-evaluation.}    
    \label{fig_rating2}
\end{figure*}

\begin{figure*}[t]
    \centering
    \includegraphics[width=\textwidth]{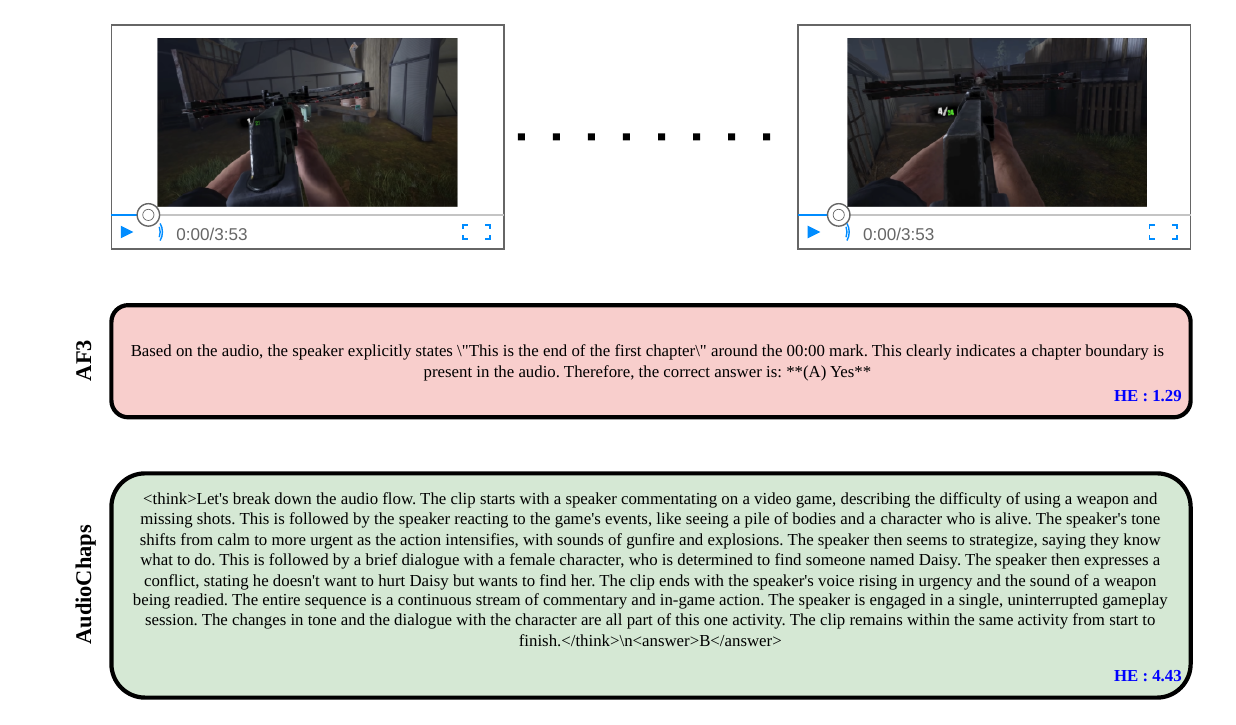}
    \caption{\small Qualitative comparison of AF3-Think-8B and AudioChaps-R1-8B on human-evaluation.}    
    \label{fig_rating3}
\end{figure*}

\begin{figure*}[t]
    \centering
    \includegraphics[width=\textwidth]{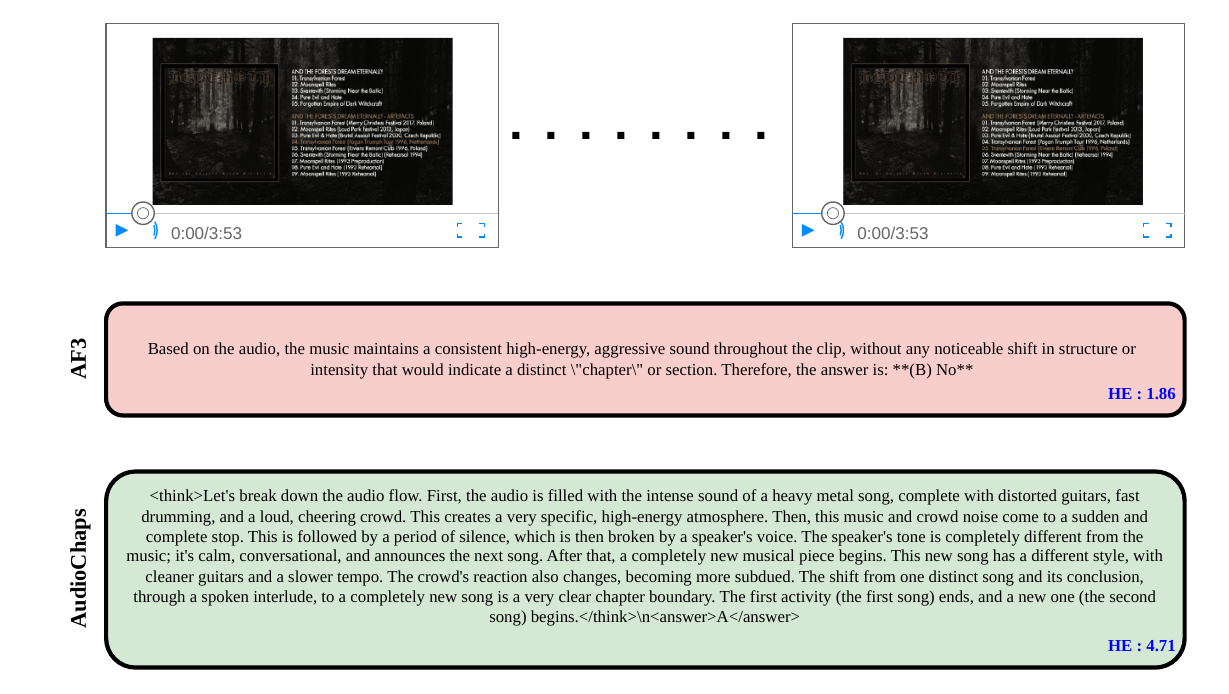}
    \caption{\small Qualitative comparison of AF3-Think-8B and AudioChaps-R1-8B on human-evaluation.}    
    \label{fig_rating4}
\end{figure*}

To assess reasoning quality beyond automatic metrics, we conducted a blind human evaluation with seven raters, all PhD students or post-doctoral researchers. For each sample, a rater was shown a 60-second audio clip and the chapter-boundary question alongside two anonymized responses, one from the AF3-Think base model and one from AudioChaps-R1, presented in randomized order to remove position bias. Each response was scored from 1 (poor) to 5 (excellent) on whether it correctly identified the presence or absence of a boundary and on the quality of its justification. Averaged over all samples and raters, AF3-Think scored 2.77 while AudioChaps-R1 scored 4.46, a clear preference for the latter's reasoning. The rater-facing instructions are reproduced in Figure~\ref{fig_eval_interface}, and representative samples with their ratings are shown in Figures~\ref{fig_rating1}, \ref{fig_rating2}, \ref{fig_rating3}, and~\ref{fig_rating4}.

\section{Additional Temporal-Localization Analysis}
\label{apx_temporal_localization}

AudioChaps is designed to align the existing reasoning capabilities of different LALM backbones with creator-authored chapter-boundary decisions. Most current LALMs do not provide sufficiently reliable timestamp predictions, so our primary task is to determine whether a chapter boundary occurs within an approximately $\pm 10$-second interval, which supports practical chapter-level navigation in long-form media. More precise temporal localization may provide additional value, but it is not the central focus of this work. We therefore report timestamp-based experiments separately as a secondary analysis for models that natively support temporal prediction.

\subsection{MOSS Timestamp Supervision}

MOSS-Think-8B has a comparatively strong native temporal-localization capability. AudioChaps can therefore exploit this capability in addition to aligning its chapter-boundary decisions. Specifically, MOSS-Think-8B can emit an estimated boundary location within the 60-second input window. For the MOSS instantiation of AudioChaps, we include a small auxiliary timestamp reward, computed using the creator-annotated boundary timestamp, alongside the boundary-classification objective. This component leverages the backbone's existing temporal capability and is not required by the primary AudioChaps framework.

The AudioChaps-adapted MOSS model substantially improves temporal localization across all four subcategories. The mean absolute error (MAE) decreases from 7.39 to 2.88 seconds on \textit{Dynamic Media}, 9.42 to 4.49 seconds on \textit{Gaming}, 5.47 to 2.46 seconds on \textit{Music}, and 6.75 to 2.98 seconds on \textit{Structured Speech}. These results show that the GRPO-based AudioChaps framework can leverage the native temporal-localization capability of MOSS-Think-8B to achieve more accurate chapter-boundary localization.

\subsection{Timestamp-Aware Recording-Level Decoding}

In the full-length audio evaluation, we additionally examine whether MOSS's in-window timestamp predictions can improve the conversion of overlapping window-level outputs into recording-level chapter boundaries. For each positive prediction from a window beginning at absolute time $t_0$, MOSS emits an in-window timestamp $t_s$. We convert this prediction into an absolute boundary vote at $t_0+t_s$. Votes are grouped using single-linkage clustering with a 10-second linkage threshold, and each cluster produces one boundary estimate at the median of its member predictions. Predicted timestamps outside the 60-second input interval are discarded.

This timestamp-based decoder is evaluated separately from the window-run decoder used in the primary full-length comparison. For MOSS-Think-AudioChaps-R1-8B, it raises precision from 48.2 to 54.3, recall from 49.3 to 64.9, and F1 from 46.2 to 55.2. It also reduces the pooled median reference-to-estimate deviation from 8 to 2 seconds. Among timestamp clusters supported by multiple overlapping windows, the median within-cluster spread is 1 second and the 90th-percentile spread is 8 seconds.

Overall, timestamp-aware decoding provides additional gains in the full-length setting when the underlying LALM supports reliable temporal outputs. These results complement the primary window-run evaluation but do not alter the central focus of AudioChaps on aligning chapter-boundary judgments with creator-authored editorial annotations.

\FloatBarrier

\begin{figure*}[t]
\centering
\begin{tcolorbox}[
    width=\textwidth,
    colback=gray!4, colframe=gray!55!black,
    title=Instructions shown to raters, fonttitle=\bfseries\small,
    boxrule=0.5pt, arc=2pt, left=4pt, right=4pt, top=3pt, bottom=3pt]
\small
\textbf{Study description.} This study evaluates an LALM trained to detect structural chapter boundaries in YouTube video clips (we consider audio only). For each sample, the models are provided with a 1-minute audio clip and a targeted prompt. You will see two randomized model responses (Output A and Output B). Please listen to the clip, review the scripts, and rate each model's ability to accurately identify whether a boundary exists, as well as the quality of its justification.

\smallskip
\textbf{Important note.} Model outputs (A, B) are randomized per sample to prevent evaluation bias. For example, Model 1's output might appear as Output A in the first sample and Output B in the next.
\end{tcolorbox}

\caption{Instructions shown to human evaluators. Each sample pairs a 1-minute clip and the boundary question with two anonymized, randomly ordered model outputs (Output A and Output B), each rated on a 1--5 scale.}
\label{fig_eval_interface}
\end{figure*}

\end{document}